\documentclass[%
 reprint,
 superscriptaddress,
 amsmath,amssymb,
 aps,
 pra,
]{revtex4-1}

\usepackage{graphicx}
\usepackage{dcolumn}
\usepackage{bm}
\usepackage{color}

\begin{document}

\preprint{APS/123-QED}

\title{Visible transitions in Ag-like and Cd-like lanthanide ions}

\author{Shunichi Murata}
\affiliation{%
Institute for Laser Science, The University of Electro-Communications, Tokyo 182-8585, Japan
}%

\author{Takayuki Nakajima}
\affiliation{%
Institute for Laser Science, The University of Electro-Communications, Tokyo 182-8585, Japan
}%

\author{Marianna S. Safronova}
\affiliation{%
University of Delaware, Newark, Delaware 19716, USA
}%
\affiliation{%
Joint Quantum Institute, NIST and the University of Maryland, Maryland 20742, USA
}%

\author{Ulyana I. Safronova}
\affiliation{%
University of Nevada, Reno, Nevada 89557, USA
}%

\author{Nobuyuki Nakamura}
\affiliation{%
Institute for Laser Science, The University of Electro-Communications, Tokyo 182-8585, Japan
}%

\date{\today}

\begin{abstract}
We present visible spectra of Ag-like ($4d^{10}4f$) and Cd-like ($4d^{10}4f^2$) ions of Ho (atomic number $Z=67$), Er (68), and Tm (69) observed with a compact electron beam ion trap.
For Ag-like ions, prominent emission corresponding to the M1 transitions between the ground state fine structure splitting $4f_{5/2}$--$4f_{7/2}$ is identified.
For Cd-like ions, several M1 transitions in the ground state configuration are identified.
The transition wavelength and the transition probability are calculated with the relativistic many-body perturbation theory and the relativistic CI + all-order approach.
Comparisons between the experiments and the calculations show good agreement.
\end{abstract}

\maketitle


\section{\label{sec:introduction}Introduction}

Visible transitions in highly charged heavy ions are of interest for many applications.
For example, transitions in highly charged W ions are in strong demand for the stable operation of the large scale fusion reactor ITER under construction.
Since W is the material of the plasma facing wall of ITER, sputtered W ions are considered to be the main impurity in the ITER plasma \cite{Skinner1}.
Thus it is important to diagnose and control the W influx and charge evolution through spectroscopic diagnostics of W ions.
Although all the wavelength ranges, including short wavelength ranges such as EUV and x-rays, are important for the diagnostics, transitions in the visible range are especially important due to the advantage that a variety of common optical components, such as mirrors, lenses, and fiber optics, can be applied.
Optical transitions in highly charged ions (HCIs) have been proposed for a new type of an optical clock that has a significantly enhanced sensitivity to the fine-structure constant variation due to the strong relativistic effects \cite{Berengut1}.
The variation of fundamental constants arises in many theories beyond the Standard Model of particle physics \cite{Uzan1} and is hinted by the astrophysical observations \cite{Webb1}.
Recently, it was suggested that dark matter may lead to oscillations of fundamental constants \cite{Arvanitaki1} or transient effects \cite{Arvanitaki1} that may be potentially detectable with such clocks \cite{Derevianko1}. 
It is also an advantage that the wavefunction of the electron tightly bound in a highly charged ion is less sensitive to the perturbation such as external fields.
A detailed assessment of systematic effects in optical clocks based on HCI was carried out in \cite{Derevianko2,Dzuba1}, reaching the conclusion that the next order of magnitude improvement in the accuracy of frequency standards to $10^{-19}$ uncertainty may also be achievable.

HCI with open 4f shell are particularly important for these applications.
For example, in the diagnostics of the ITER plasma, it has been pointed out that the diagnostics and control of the edge plasma are extremely important for the steady state operation of high-temperature plasmas \cite{Mita1,Clementson4}.
The impurity in the edge region will be dominated by W ions with charge states below Ag-like ($4d^{10}4f$) \cite{Biedermann6}; thus the spectroscopic diagnostics of $4f$ open shell W ions plays a key role in the operation of ITER.
For the application to an atomic clock, $4f$--$5s$ transitions in medium atomic number ($Z$) ions are potential candidates for a clock transition because they can fall in the optical range due to their nearly degenerate energy levels \cite{Berengut1}.
For example, Ag-like ($4d^{10}4f$ or $4d^{10}5s$) to Sn-like ($5s^24f^2$ or $5s^25p^2$) lanthanide (Ce, Pr, Nd, and Sm) ions have been proposed by Safronova et al. \cite{Safronova3}, and I-like Ho ($4d^{10}4f^65s$) has been proposed by Dzuba et al. \cite{Dzuba1}

In order to meet these demands, visible transitions in $4f$ open shell mid-$Z$ ions have been extensively studied recently both experimentally and theoretically.
For example, visible spectra of W ions have been extensively studied with electron beam ion traps (EBITs) \cite{Qiu2,Komatsu1}.
Theoretical analysis for identifying those experimental spectra has also been performed \cite{Ding2,Safronova5}.
For the visible transitions useful for an atomic clock, theoretical studies \cite{Berengut1,Safronova3,Dzuba1} have taken a lead role, and several experimental efforts \cite{Nakajima1,Windberger1} to identify the proposed transitions have followed the theoretical studies.
In spite of such efforts, complex fine structure levels arising from the $4f$ open shell often hamper the detailed understandings of the experimental spectra.
Indeed, identification of the Ho$^{14+}$ lines observed with a compact EBIT has not yet been achieved due to its complex structure arising from the $4d^{10}4f^65s$ and $4d^{10}4f^55s^2$ configurations \cite{Nakajima1}.
Theoretically, precision calculations even of simpler systems with 3-4 electrons in the valence 4f shell are very challenging, with a very few available experiential benchmarks.
No precision method currently exists to accurately treat ions with the mid-filled $4f$ shell.

In this paper, we study the simplest $4f$ open shell systems, i.e. Ag-like ($4d^{10}4f$) and Cd-like ($4d^{10}4f^2$) configurations, as the first step towards systematic understanding of complex $4f$ open shell systems.
$Z$ dependence also provides important information; thus we study three lanthanide ions with $Z=67$ (Ho), 68 (Er), and 69 (Tm).
We present visible spectra of these ions obtained using a compact EBIT.
We also evaluate the atomic properties of these ions using the relativistic many-body perturbation theory (RMBPT) code and the CI+all-order (SD) approach.
The level energies, transition wavelengths, and transition rates are evaluated.
For Ag-like ions, the fine structure splitting between $4f_{7/2}$ and $4f_{5/2}$ is experimentally determined from the direct measurement of the M1 transition between the fine structure levels, and compared with the results obtained by RMBPT, as well as the existing experimental and theoretical values.
For Cd-like ions,
we compare the energy levels evaluated by the CI+all-order (SD) approach. with results obtained by the RMBPT code.
Experimental transition wavelengths in several Cd-like lines are compared with the results obtained by the CI+all-order (SD) approach.

\section{\label{sec:exp}Experiments}

The present experiments were performed using a compact EBIT, called CoBIT.
Detailed description of CoBIT is given in Ref.~\cite{cobit}.
Briefly, it consists of an electron gun, a drift tube (DT), an electron collector, and a high-critical-temperature superconducting magnet.
The DT is composed of three successive cylindrical electrodes that act as an ion trap by applying a positive trapping potential (typically 30~V) at both ends (DT1 and 3) with respect to the middle electrode (DT2).
The electron beam emitted from the electron gun is accelerated towards the DT while it is compressed by the axial magnetic field (typically 0.08~T) produced by the magnet surrounding the DT.
The space charge potential of the compressed high-density electron beam acts as a radial trap in combination with the axial magnetic field.
Highly charged ions are produced through the successive ionization of the trapped ions.
The lanthanide element of interest was introduced with an effusion cell \cite{yamada1}.
The temperature of the cell was 900 - 950~$^\circ$C.

The setup for observing visible spectra was essentially the same as that used in our previous studies, where the visible emission lines of tungsten ions were observed \cite{Komatsu1,Komatsu2,Kobayashi2}.
Briefly, the emission from the trapped lanthanide ions was focused by a convex lens on the entrance slit of a commercial Czerny-Turner type of visible spectrometer (Jobin Yvon HR-320) with a focal length of 320 mm.
For survey observations, a 300~gr/mm grating blazed at 500~nm was used, whereas for wavelength determination, a high resolution 1200~gr/mm grating blazed at 400~nm was used.
The diffracted light was detected with either a liquid-nitrogen-cooled back-illuminated CCD (Princeton Instruments Spec-10:400B LN) operated at -115~$^\circ$C or a Peltier-cooled back illuminated CCD (Andor iDus 416) operated at -70~$^\circ$C.
The wavelength scale was calibrated using emission lines from several standard lamps placed outside CoBIT.
The uncertainty in the experimentally determined wavelength was estimated from reproducibility to be about 0.06~nm.


The electron beam energy determines the maximum charge state which can be produced in the trap according to the ionization energy listed in Table~\ref{tab:ip}.
For example, for Ho, when the electron energy is below 439~eV, the maximum charge state is In-like and Cd-like can not be produced.
In other words, the emission line from Cd-like Ho can be observed only when the electron energy exceeds 439~eV.
Thus the charge state of the ion that should be assigned to each observed line can be determined by observing the electron energy dependence.


\begin{table}[tb]
\caption{\label{tab:ip}%
Ionization energy of the ions of interest in eV \cite{Scofield0}.
For example, ``In$\rightarrow$Cd" represents the ionization energy of the In-like ion.
}
\begin{ruledtabular}
\begin{tabular}{ccccc}
element&Sn$\rightarrow$In&In$\rightarrow$Cd&Cd$\rightarrow$Ag&Ag$\rightarrow$Pd\\
\colrule
Ho&400&439&478&519\\
Er&443&483&524&566\\
Tm&488&529&571&615\\
\end{tabular}
\end{ruledtabular}
\end{table}

\section{\label{sec:theory}Calculations}



The energy levels of Ag-like Ho$^{20+}$, Er$^{21+}$, and Tm$^{22+}$ have been obtained using the second-order RMBPT method, described in Ref.~\cite{Safronova6}, where the second and third order many-body calculations of energies and matrix elements were carried out for ions of the Ag-like sequence.
In Ref.~\cite{Safronova6}, second and third-order corrections to energies and dipole matrix elements were included for neutral Ag and for ions with $Z<60$, and second-order corrections were included for $Z>60$.
The effect of third order is small for the M1 transitions in HCI.

\begin{table*}
\caption{\label{tab:en} Energies in cm$^{-1}$ for the lowest levels in Cd-like Ho, Er, and
Tm ions evaluated by the RMBPT and CI+all-order methods.}
\begin{ruledtabular}
\begin{tabular}{crrrrrrrrr}
\multicolumn{1}{c}{Level}&
\multicolumn{2}{c}{Cd-like Ho}&
\multicolumn{2}{c}{Cd-like Er}&
\multicolumn{2}{c}{Cd-like Tm}\\
\multicolumn{1}{c}{}&
\multicolumn{1}{c}{CI+all}&
\multicolumn{1}{c}{RMBPT} &
\multicolumn{1}{c}{CI+all}&
\multicolumn{1}{c}{RMBPT} &
\multicolumn{1}{c}{CI+all}&
\multicolumn{1}{c}{RMBPT} \\
\hline \\[-0.6pc]
 $^3\!H_{4}$ &  0&  0&  0&  0& 0 &  0\\
 $^3\!H_{5}$ & 11001& 11164 & 12591 & 12779& 14356 & 14540\\
 $^3\!F_{2}$ & 12439& 12642 & 13256 & 13411& 13983 & 14191\\
 $^3\!F_{3}$ & 20362& 20626 & 22417 & 22662& 24565 & 24843\\
 $^1\!G_{4}$ & 20872& 20945 & 22887 & 22933& 24980 & 25060\\
 $^3\!H_{6}$ & 21288& 21581 & 24210 & 24535& 27407 & 27729\\
 $^3\!F_{4}$ & 34539& 34714 & 38361 & 38543& 42449 & 42663\\
 $^1\!D_{2}$ & 44630& 44850 & 47795 & 47910& 50868 & 51096\\
 $^3\!P_{0}$ & 51158& 51511 & 54085 & 54274& 56731 & 57082\\
 $^3\!P_{1}$ & 55386& 55823 & 59060 & 59387& 62638 & 63086\\
 $^1\!I_{6}$ & 56444& 55863 & 60316 & 59702& 64302 & 63732\\
 $^3\!P_{2}$ & 62787& 63270 & 67713 & 68120& 72752 & 73246\\
 $^1\!S_{0}$ &120738& 120395 &128181& 127515&135190& 134824\\
\end{tabular}
\end{ruledtabular}
\end{table*}

The CI+all-order (SD) approach described in Ref.~\cite{Safronova5} was applied in this study for calculating the energies, wavelengths, and transition rates of low-lying levels in Cd-like Ho$^{19+}$, Er$^{20+}$, and Tm$^{21+}$ ions.
Table~\ref{tab:en} shows the energies for all the fine structure levels of $4f^2$ configuration.
The values obtained with the RMBPT code are also listed in the table for comparison.
The smallest (0.20-0.35\%) difference for energies evaluated by CI+all-order and RMBPT codes is for the energies for the $^1\!G_4$ level.
The largest (1.46-1.47\%) difference for energies evaluated by CI+all-order and RMBPT codes is for the energies for the $^3\!H_5$ level.

\begin{figure}[tb]
\includegraphics[width=0.45\textwidth]{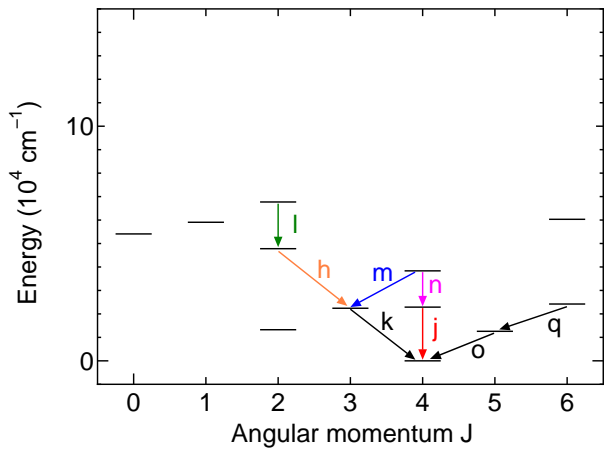}
\caption{\label{fig:cd_level}
Fine structure levels of the $4f^2$ ground state configuration in Cd-like Er calculated with the CI+all-order code.
The solid arrows represent the transitions observed in the present experiment.
The labels on the arrows are listed in Table~\ref{tab:rate}.
}
\end{figure}


Figure~\ref{fig:cd_level} shows the energy levels of Cd-like Er obtained by the CI+all-order code.
The energy level structure for Cd-like Ho and Tm is practically the same as that for Cd-like Er, i.e. the level ordering is not changed in this $Z$ region except for the order between $^3\!H_{5}$ and $^3\!F_{2}$ (see Table~\ref{tab:en}).

\section{\label{sec:results}Results and discussion}


\begin{figure}[t]
\includegraphics[width=0.45\textwidth]{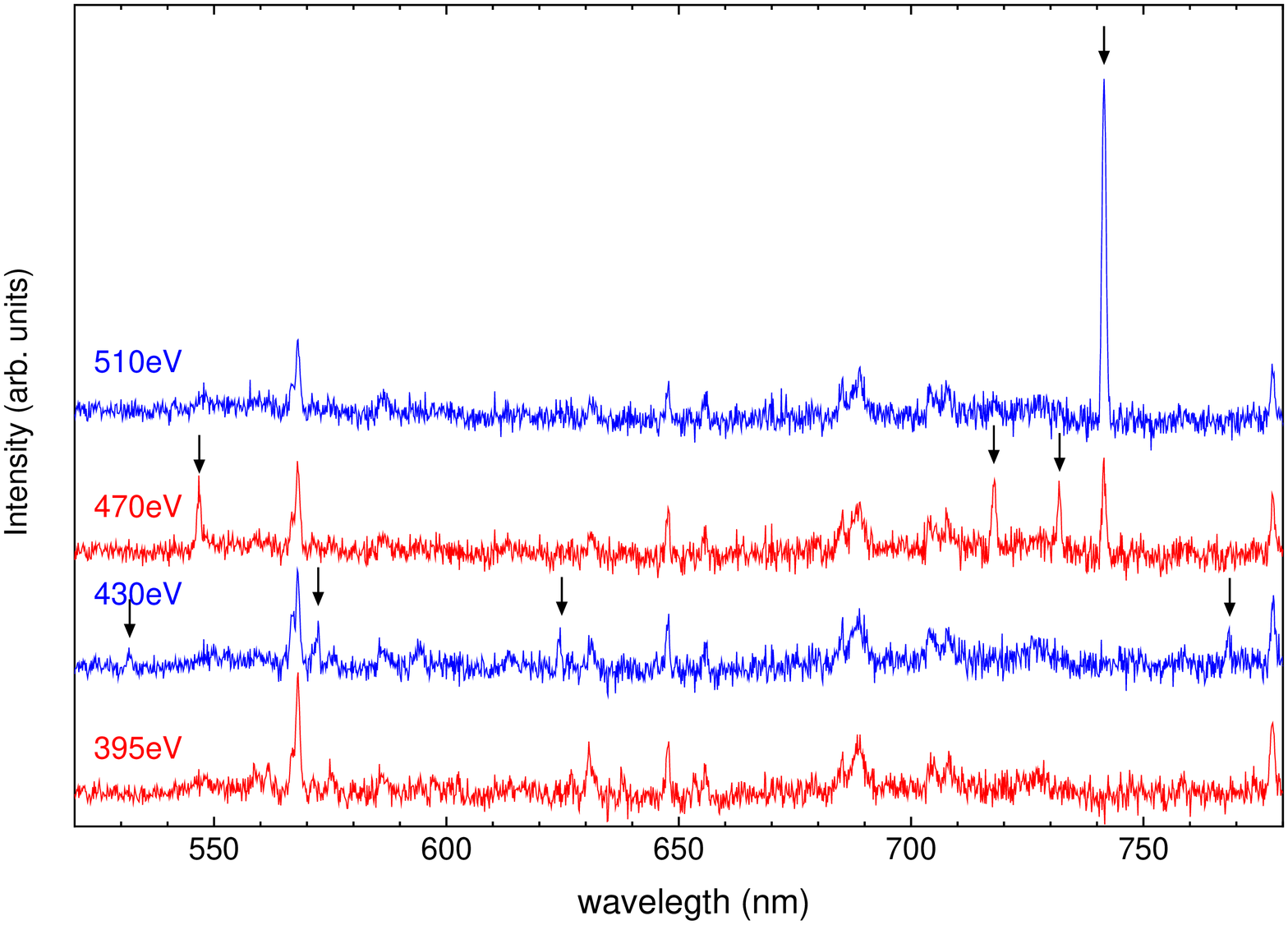}
\caption{\label{fig:edep}
Electron beam energy dependence of visible spectra of Ho ions.
}
\end{figure}
Figure \ref{fig:edep} shows the electron energy dependence of survey spectra of Ho.
According to the ionization energy values (Table~\ref{tab:ip}), In-like Ho can not be produced with an electron energy of below 400~eV; thus the spectrum obtained at 395~eV should not contain any lines from In-like Ho.
When the electron energy was increased to 430~eV, i.e. above the threshold energy for producing In-like Ho, the lines indicated by arrows appeared; thus they can be assigned to In-like Ho.
Similarly, the lines indicated by the arrows in the 470~eV spectra can be assigned to Cd-like Ho because they appeared when the electron energy exceeded the threshold energy for producing Cd-like Ho.
The line at around 740~nm also appeared at an electron energy of 470~eV similarly with the Cd-like lines.
However, when the energy was further increased, the line at 740~nm kept increasing and became most prominent at 510~eV whereas the lines assigned to Cd-like decreased and almost disappeared at 510~eV.
Thus we have assigned the prominent line at 740~nm to Ag-like Ho.
The wavelength is consistent with the $^2\!F_{7/2}$-$^2\!F_{5/2}$ fine structure splitting value 13500~cm$^{-1}$ derived from the interpolation of the experimentally determined fine structure splitting of other elements \cite{Sugar3}.
We thus believe that the present assignment, but have no idea at present why the Ag-like line was observed below the threshold energy.
Appearance of a line below the threshold has often been observed in spectroscopy with an EBIT \cite{Sakoda1,Windberger2}.
For example, in our previous study \cite{Sakoda1}, a Rh-like Ba$^{11+}$ line was observed below the threshold energy.
It can be understood by considering the ionization from the $4d^95s$ metastable state of Ba$^{10+}$ as discussed in Ref.~\cite{Sakoda1}.
However, it is unlikely in the present case because there is no metastable state in the electronic excited states of the Cd-like system.
The energy dependence of the Er and Tm spectra was also studied and the charge state was assigned to the observed lines similarly to the Ho spectra.

\begin{figure}[t]
\includegraphics[width=0.45\textwidth]{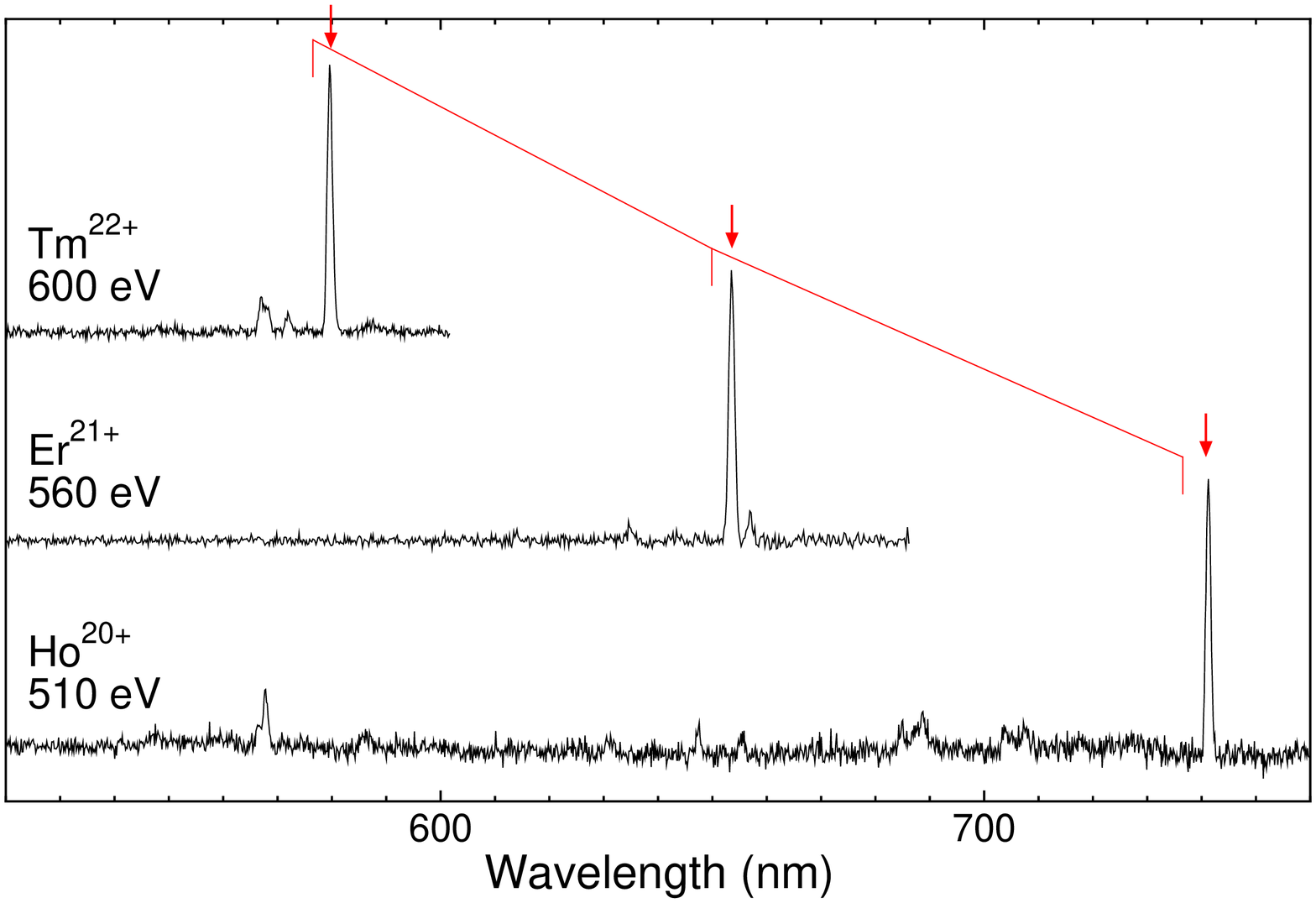}
\caption{\label{fig:ag}
Visible spectra of Ag-like Ho$^{20+}$, Er$^{21+}$, and Tm$^{22+}$.
The prominent line indicated by the arrows in each spectrum is the $^2\!F_{7/2}$-$^2\!F_{5/2}$ transition.
The vertical bars connected by the solid line represent the theoretical wavelength calculated with RMBPT.
Note that the intensity can not be compared.
}
\end{figure}

Figure~\ref{fig:ag} shows spectra of Ho, Er, and Tm obtained at the electron energies where the dominance of the Ag-like line was confirmed (510, 560, and 600~eV, respectively).
The line indicated by the arrow in each spectrum correspond to the M1 transitions between the ground state fine structure splitting $4f_{5/2}$--$4f_{7/2}$.
The vertical bars connected with a solid line represent the theoretical wavelength obtained in the present RMBPT calculations.
As seen in the figure, the theoretical values are in good agreement with the experimental ones although there can be found a slight shift between them.
The transition was also observed with a high resolution 1200~gr/mm grating for wavelength determination.
The fine structure splitting energies derived from the experimental wavelengths are listed in Table~\ref{tab:ag} together with the available experimental and theoretical values.
The multi-configuration Dirac-Fock (MCDF) values \cite{Ding2} were obtained by extensive relativistic configuration interaction calculations using the GRASP2K package.
The active space techniques have been employed to extend the configuration expansion systematically.
The electron correlation effects, Breit interaction and quantum electrodynamics effects to the atomic state wavefunctions and the corresponding energies had been taken into account.
Multi-configuration Dirac-Hartree-Fock (MCDHF) method was also employed in Ref.~\cite{Grumer1}.
The work in Ref.~\cite{Ivanova2} used relativistic perturbation theory with a zero-approximation model potential (RPTMP).
As seen in the table, the theoretical values are in good agreement with the present experimental values.

It is noted that the experimental values by Zhao \cite{Zhao2} and Sugar \cite{Sugar3} were not directly measured, but determined from the experimental values for other $Z$.
Thus the present result is the first direct determination of the fine structure levels.
As seen in the table, the present experimental values agree with existing experimental and theoretical values.

\begin{table*}
\caption{\label{tab:ag}
Fine structure splitting $4f_{5/2}$--$4f_{7/2}$ in Ag-like Ho, Er, and Tm (cm$^{-1}$).
}
\begin{ruledtabular}
\begin{tabular}{cccccccc}
&\multicolumn{3}{c}{exp}&\multicolumn{4}{c}{theory}\\
\cline{2-4} \cline{5-8}
element($Z$)&present&Zhao\footnote{$Z$-scaled value derived from the experimental fine structure splitting for several elements between $Z=62$ and 74 \cite{Sugar3,Zhao2}.}&
Sugar\footnote{Linearly interpolated value derived from the experimental fine structure splitting for other $Z$ \cite{Sugar3}.}&
RMBPT&
MCDF\cite{Ding2}&RPTMP\cite{Ivanova2}&MCDHF\cite{Grumer1}\\ \hline
Ho (67)&13487(2)&13392&13500&13573&
13140&13486&13509\\
Er (68)&15297(2)&15239&&15383&
14926&15371&15320\\
Tm (69)&17258(2)&17271&&17341&
16871&17695&17280\\
\end{tabular}
\end{ruledtabular}
\end{table*}

\begin{figure}
\includegraphics[width=0.45\textwidth]{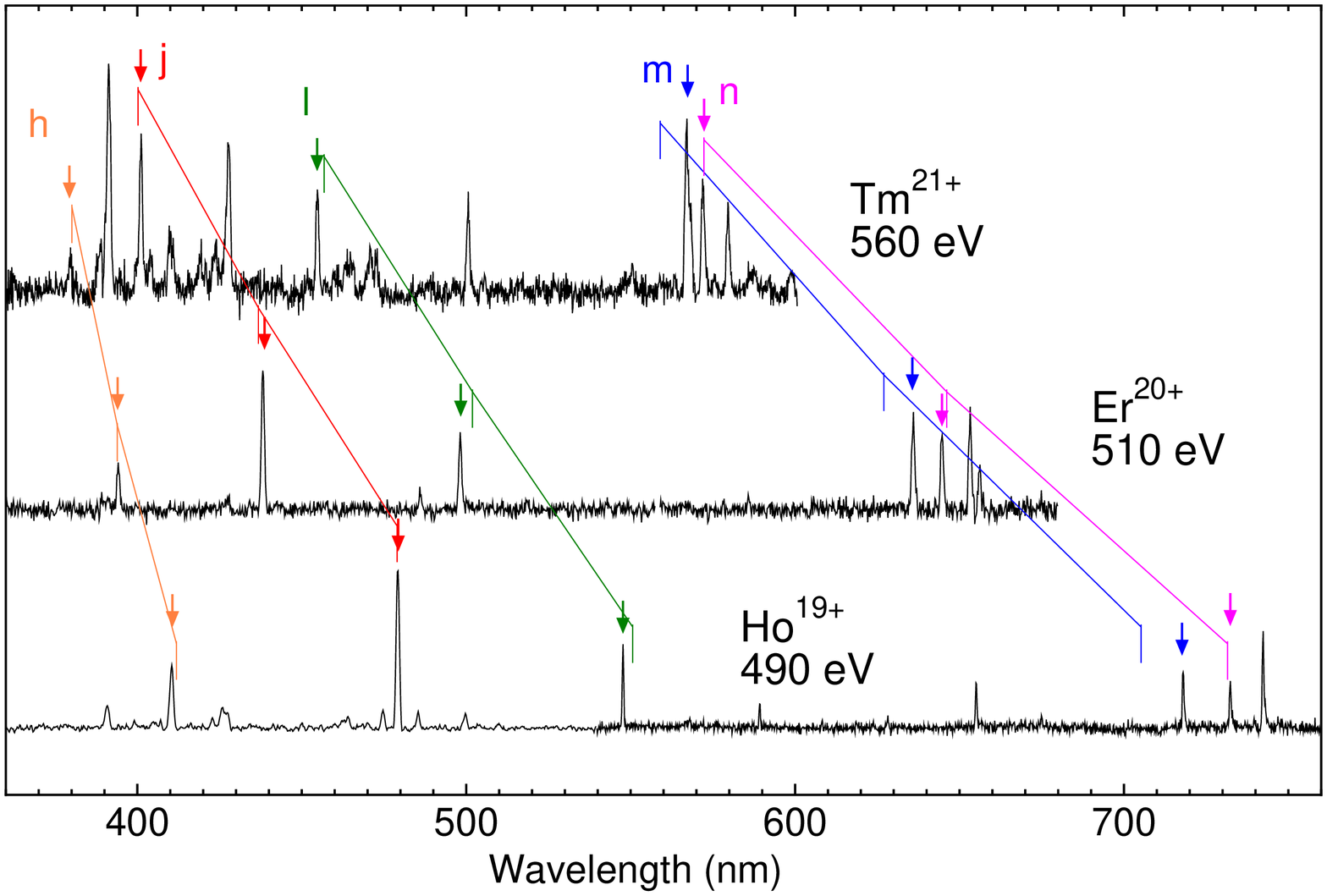}
\caption{\label{fig:cd}
Visible spectra of Cd-like Ho, Er, and Tm.
The lines assigned to Cd-like from the energy dependence measurements are indicated by the arrows.
The other lines are originated from other charge states or impurity ions.
Note that the intensity can not be compared.
The vertical bars connected by the solid line represent the CI+all-order wavelengths for five prominent lines.
}
\end{figure}

\begin{table*}
\caption{\label{tab:rate} Wavelength (nm in air) and transition rates for the lowest levels in Cd-like Ho, Er, and
Tm ions evaluated by the CI+all-order method.
}
\begin{ruledtabular}
\begin{tabular}{ccrrrrrrrrr}
&&
\multicolumn{3}{c}{Cd-like Ho}&
\multicolumn{3}{c}{Cd-like Er}&
\multicolumn{3}{c}{Cd-like Tm}\\
\cline{3-5} \cline{6-8} \cline{9-11}
&&\multicolumn{2}{c}{wavelength}&&\multicolumn{2}{c}{wavelength}&&\multicolumn{2}{c}{wavelength}&\\
\cline{3-4} \cline{6-7} \cline{9-10}
\multicolumn{1}{c}{Label}&
\multicolumn{1}{c}{Term}&
\multicolumn{1}{c}{exp.}&
\multicolumn{1}{c}{theory}&
\multicolumn{1}{c}{$gA_r$}&
\multicolumn{1}{c}{exp.}&
\multicolumn{1}{c}{theory}&
\multicolumn{1}{c}{$gA_r$}&
\multicolumn{1}{c}{exp.}&
\multicolumn{1}{c}{theory}&
\multicolumn{1}{c}{$gA_r$}\\
\hline \\[-0.6pc]
a& $^3\!F_{2}-\ ^3\!P_{2} $&      & 198.6 & 6.65[0]&      & 183.6 & 9.80[0]&      & 170.1& 1.38[1]\\
b& $^3\!H_{5}-\ ^1\!I_{6} $&      & 220.0 & 4.61[2]&      & 209.5 & 6.55[2]&      & 200.2& 9.26[2]\\
c& $^3\!F_{2}-\ ^3\!P_{1} $&      & 232.8 & 1.06[1]&      & 218.3 & 1.59[1]&      & 205.5& 2.39[1]\\
d& $^3\!F_{3}-\ ^3\!P_{2} $&      & 235.6 & 1.02[2]&      & 220.7 & 1.50[2]&      & 207.5& 2.18[2]\\
e& $^3\!H_{6}-\ ^1\!I_{6} $&      & 284.4 & 2.96[2]&      & 276.9 & 3.91[2]&      & 271.0& 5.12[2]\\
f& $^3\!H_{4}-\ ^3\!F_{4} $&      & 289.4 & 4.79[1]&      & 260.6 & 6.11[1]&      & 235.5& 7.61[1]\\
g& $^3\!F_{2}-\ ^1\!D_{2} $&      & 310.6 & 2.29[2]&      & 289.4 & 3.16[2]&      & 271.0& 4.34[2]\\
h& $^3\!F_{3}-\ ^1\!D_{2} $&410.82& 411.9 & 1.67[2]&394.21& 393.9 & 2.10[2]&379.25& 380.1& 2.59[2]\\
i& $^3\!H_{5}-\ ^3\!F_{4} $&      & 424.7 & 7.79[1]&      & 387.9 & 1.05[2]&      & 355.9& 1.41[2]\\
j& $^3\!H_{4}-\ ^1\!G_{4} $&479.88& 479.0 & 1.45[2]&438.21& 436.8 & 2.09[2]&400.91& 400.2& 2.99[2]\\
k& $^3\!H_{4}-\ ^3\!F_{3} $&485.98& 491.0 & 7.19[0]&442.34& 446.0 & 1.08[1]&
& 407.0& 1.61[1]\\
l& $^1\!D_{2}-\ ^3\!P_{2} $&546.83& 550.6 & 2.19[2]&498.29& 501.9 & 3.09[2]&454.50& 456.8& 4.36[2]\\
m& $^3\!F_{3}-\ ^3\!F_{4} $&718.00& 705.2 & 2.53[2]&636.35& 627.0 & 3.67[2]&566.67& 559.0& 5.27[2]\\
n& $^1\!G_{4}-\ ^3\!F_{4} $&731.77& 731.5 & 2.11[2]&645.07& 646.1 & 3.08[2]&571.74& 572.3& 4.45[2]\\
o& $^3\!H_{4}-\ ^3\!H_{5} $&      & 908.8 & 3.54[2]&791.26& 794.0 & 5.27[2]&
& 696.4& 7.75[2]\\
p& $^1\!D_{2}-\ ^3\!P_{1} $&      & 929.5 & 1.82[1]&      & 887.5 & 2.27[1]&      & 849.4& 2.81[1]\\
q& $^3\!H_{5}-\ ^3\!H_{6} $&      & 971.8 & 3.09[2]&860.89& 860.4 & 4.43[2]&
& 766.0& 6.25[2]\\
r& $^3\!H_{5}-\ ^1\!G_{4} $&      &1012.8 & 1.55[1]&      & 971.0 & 1.94[1]&      & 941.0& 2.35[1]\\
s& $^3\!F_{2}-\ ^3\!F_{3} $&      &1261.8 & 8.19[1]&      &1091.3 & 1.26[2]&      & 944.7& 1.91[2]\\
t& $^3\!P_{1}-\ ^3\!P_{2} $&      &1350.8 & 2.10[1]&      &1155.4 & 3.28[1]&      & 988.5& 5.09[1]\\
\end{tabular}
\end{ruledtabular}
\end{table*}

Figure~\ref{fig:cd} shows spectra of Ho, Er, and Tm obtained at the electron energies where the dominance of the Cd-like line was confirmed (490, 510, and 560~eV, respectively).
The lines indicated by the arrows represent the transitions assigned to Cd-like from the energy dependence.
The vertical bars connected with a solid line represent the theoretical wavelengths for the five prominent lines obtained in the present RMBPT calculations.
The transition wavelengths obtained in high resolution measurements are listed in Table~\ref{tab:rate} together with the theoretical wavelengths and transition rates obtained with the CI+all-order approximation.
The M1 and E2 transition rates were evaluated based on Eqs. 1 and 2 in Ref.~\cite{Safronova5}.
It was demonstrated in Ref.~\cite{Safronova5} that M1 transition rates are larger by factor of 10 - 100 than E2 contributions.
As a result, we present in Table~\ref{tab:rate},the M1 transition rate for 20 transitions in Ho$^{19+}$, Er$^{20+}$, and Tm$^{21+}$ ions.
The $gA_r$ values are charged from 10~s$^{-1}$ up 1000~s$^{-1}$.
Our CI+all-order results are in good agreement with experimental results as confirmed in Fig.~\ref{fig:cd} and Table~\ref{tab:rate}.


According to the energy diagram shown in Fig.~\ref{fig:cd_level}, the sum of the transition energies of m and k should be equal to that of n and j.
Although the line k is rather weak, it was assigned in the high resolution spectrum for Ho and Er.
The sum of the observed transition energies (converted to the vacuum values) is confirmed to be 34495~cm$^{-1}$ for Ho and 38311~cm$^{-1}$ for Er, for both m+k and n+j.
This confirms the reliability of the present assignment.

\vspace{-3mm}
\section*{Acknowledgment}
This work was mainly supported by JSPS KAKENHI Grant Number JP16H04028.
NN thanks Prof. H. Tanuma (TMU) for lending the CCD used in the visible spectroscopy.
MSS acknowledges the support of the NSF grant PHY-1620687 (USA).

\bibliography{ref}

\end{document}